\RequirePackage[OT1]{fontenc}
\documentclass[preprint,12pt]{elsarticle}

\usepackage{tabularx,ragged2e,booktabs}
\usepackage{enumitem}
\usepackage{amsmath}

\usepackage{amstext}
\usepackage{array}
\usepackage{ltablex}
\usepackage{caption}
\usepackage{multirow}
\usepackage{rotating}
\usepackage{subcaption}
\usepackage{graphicx}
\usepackage{adjustbox}
\usepackage{chngpage}
\usepackage{enumitem}
\usepackage{etoolbox}
\usepackage{hyperref}
\usepackage{cleveref}
\usepackage{mathrsfs}
\usepackage{bm}
\usepackage{eucal}
\usepackage{csquotes}
\usepackage{tabularx}
\usepackage{multirow}
\usepackage{hyphenat}
\usepackage{float}

\newcolumntype{s}{>{\centering\arraybackslash}>{\hsize=.1\hsize}X}

\MakeOuterQuote{"}

\newcolumntype{P}[1]{>{\RaggedRight}p{#1}}
\newcolumntype{C}[1]{>{\Centering}p{#1}}
\newtheorem{definition}{Definition}

\journal{Online Social Networks and Media}

\begin{document}

\begin{frontmatter}



\title{Social Search: retrieving information in Online Social Platforms - A Survey}


\author[inst1,inst2]{Maddalena Amendola}

\affiliation[inst1]{organization={Department of Computer Science},
            addressline={ University of Pisa}, 
            city={Pisa},
            postcode={56127},
            country={Italy}}

\author[inst2]{Andrea Passarella}

\affiliation[inst2]{organization={Institute for Informatics and Telematics (IIT)},
            addressline={ National Research Council (CNR)}, 
            city={Pisa},
            postcode={56124},
            country={Italy}}
            
\author[inst3]{Raffaele Perego}
\affiliation[inst3]{organization={Institute of Information Science and Technologies (ISTI)},
            addressline={National Research Council (CNR)}, 
            city={Pisa},
            postcode={56124},
            country={Italy}}

\begin{abstract}
\textit{Social Search} research studies methodologies exploiting social information to better satisfy user information needs in Online Social Media while simplifying the search effort and consequently reducing the time spent and the computational resources utilized. Starting from previous studies, in this work, we analyze the current state of the art of the Social Search area, proposing a new taxonomy and highlighting current limitations and open research directions. We divide the Social Search area into three subcategories, where the social aspect plays a pivotal role: \textit{Social Question\&Answering}, \textit{Social Content Search}, and \textit{Social Collaborative Search}. For each subcategory, we present the key concepts and selected representative approaches in the literature in greater detail.
We found that, up to now, a large body of studies model users' preferences and their relations by simply combining social features made available by social platforms. It paves the way for significant research to exploit more structured information about users' social profiles and behaviours (as they can be inferred from data available on social platforms) to optimize their information needs further.
\end{abstract}



\end{frontmatter}
\newpage
\section{Introduction}
Online Social Network (OSN) platforms have been one of the major evolutions of the Web for at least a decade. Aside from their social implications, OSNs are impacting many established areas of computer science, which are also progressing based on the pervasive diffusion of OSN services among users. A chief example is the area of \textit{Social Search}. Social Search lies at the intersection between Information Retrieval (IR) and Social Networks. It is generally defined as a cooperative process that relies on implicit or explicit information about social relationships between users to satisfy an information need. The main focus of this paper is to provide a comprehensive survey of the main results in this very active field. IR processes are today at the basis of almost all the most common uses of the Web, whether the user searches for a restaurant, the solution to a problem, wants to learn something, or asks for opinions. With the progressive development of tools for creating and sharing personal content and the increase of explicit or hidden interactions between users, the process that leverages users' social information in the standard IR tasks is becoming of fundamental importance. Furthermore, the interplay between IR and OSNs is becoming increasingly crucial as OSNs are one of the leading platforms where users search for information. Therefore, social content in OSNs can complement traditional Web sources in providing fresh and almost real-time information about events and ongoing discussions \cite{teevan2011twittersearch}. The reason why people prefer, in many cases, to retrieve information from a social network rather than a conventional search engine lies in the origin of the information need. Through social networks, users can obtain tailored information and different opinions by interacting with the virtual counterpart of real individuals \cite{jeon2013value}.

\begin{figure}[!ht]
    \centerline{\includegraphics[width=1\textwidth]{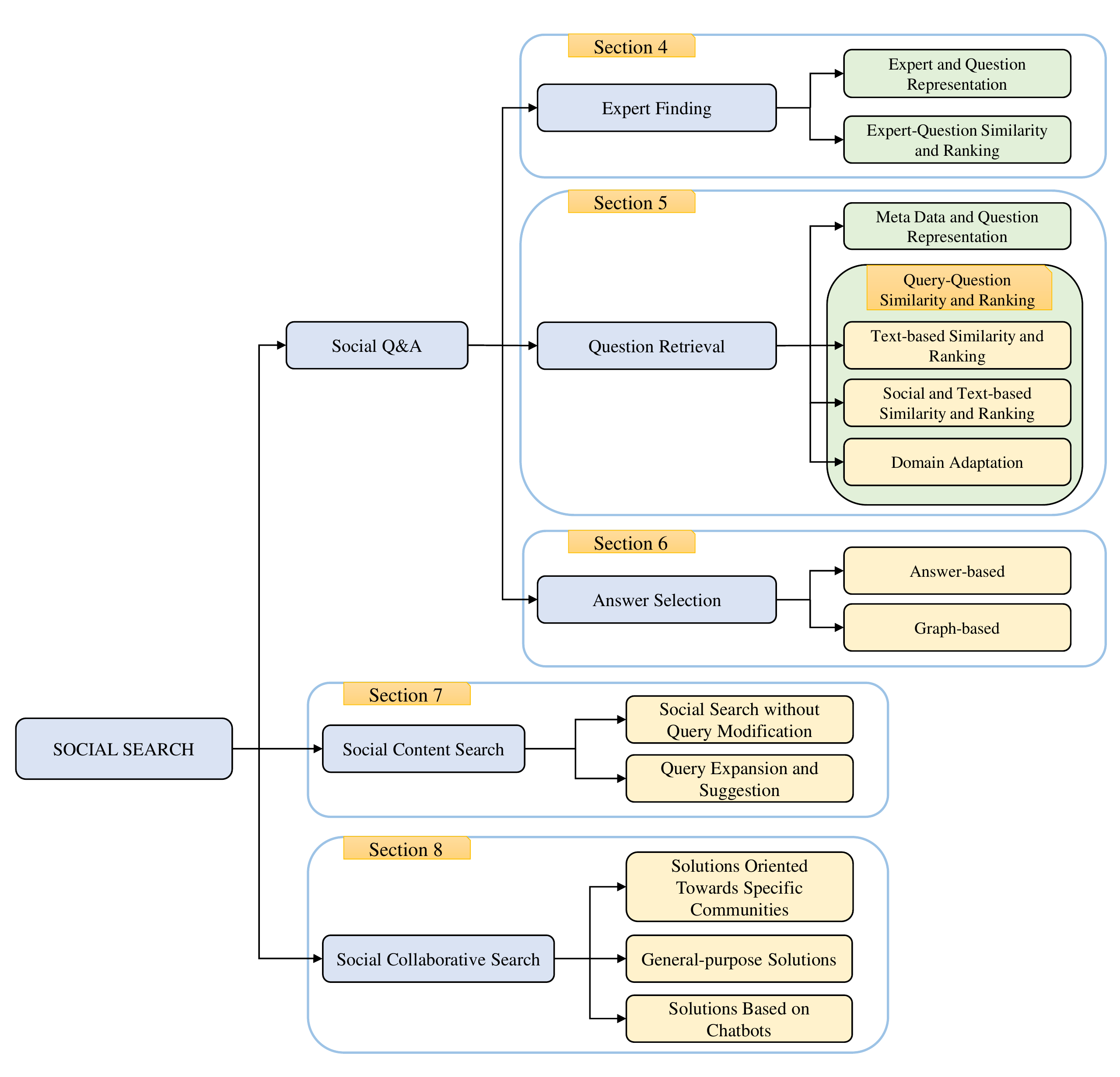}}
    \caption{The proposed Social Search taxonomy.}
    \label{fig:taxonomy}
\end{figure}

In this survey, we follow the definition and taxonomy proposed in \cite{bouadjenek2016social} for Social Search (we refer the reader to Section~\ref{sec:related} for related surveys and definitions). Thus, we consider Social Search as the IR process applied to social network platforms and content. In presenting the recent advances in Social Search, as in~\cite{bouadjenek2016social}, we divide the broad topic into three categories based on the type of Social Search system. Specifically, the following three categories can be seen as a particular instance of a Social Search system focusing on a specific approach. Figure \ref{fig:taxonomy} shows the taxonomy we propose. The leaves in the taxonomy have different colours based on their role in the specific task: green if they are components needed to perform the task, and yellow if they are alternative approaches to accomplish the task.
The three subcategories are the following:
\begin{itemize}
    \item \textbf{Social Q\&A}: through Question\&Answering (Q\&A) communities, people can ask questions using more "natural" and flexible forms of natural language (be it via speech interfaces or typing text concerning more rigid keyword-based queries), expressing their information need more completely. Alternatively, the user can refer to existing questions and avoid waiting for someone to answer. These platforms differ from social networks in the type of relationships between users, which is usually implicit here. By analyzing them, we have divided this category into three subcategories: \textit{Expert Finding}, \textit{Question Retrieval}, and \textit{Answer Selection}. The Expert Finding problem aims to identify experts in the community to whom forward a newly posted question to satisfy the user information need in the shortest possible time. The Question Retrieval problem analyses the question archives to find old questions similar to the new ones published. Finally, Answer Selection deals with the problem of automatically selecting the best answer in a long thread of questions' answers and comments, which makes users unable to choose the correct one. Expert Finding, Question Retrieval, and Answer Selection are three very widely studied problems, which are therefore analyzed separately in Sections \ref{sec:expert_finding}, \ref{sec:question_retrieval}, and \ref{sec:answer_selection}, respectively. These three topics should be seen as complementary components of a fully-fledged Social Q\&A system rather than as alternative ways to implement such a system;
    
    \item \textbf{Social Content Search}: refers to techniques that facilitate the users' searches in social networks, where the relationships between users are usually explicit and pre-established, unlike community Question\&Answering (cQA) sites where they are usually implicit. In a Q\&A platform, a user may also have no relationship with other users and still exploit the system by asking questions and taking advantage of the knowledge provided by "strangers." Instead, users usually use social networks to post/search for statuses, get information on events, stay up to date, and maintain connections. We divide the Social  Content Search literature into two areas representing two alternative approaches to accomplish the task: the works that do not modify the query and those that propose terms for query expansion or suggest queries. The literature on Social Content Search is presented in Section \ref{sec:content_search};
    
    \item \textbf{Social Collaborative Search}: in Social Q\&A and Social Content Search, the collaboration among users in the IR process is mostly implicit. However, in other cases, the search process involves multiple people who \emph{explicitly} collaborate to achieve a common purpose. Collaborative Search allows for unifying knowledge, skills, and experience. We divide this area into three parts: (i) specific community-oriented solutions (i.e., academic purposes), (ii) general-purpose solutions, and (iii) chatbot-based solutions. The first two approaches are alternatives that define the target of the Collaborative system. The last is a complementary organization of the Collaborative system that favours building dynamic communities focused on a common goal. The Social Collaborative Search literature is presented in Section \ref{sec:collaborative_search}.
\end{itemize}
Note that, to keep the paper self-contained despite the vastity of the literature, we use the following approach unless otherwise stated. We identify critical issues addressed in the literature for each leaf node in the taxonomy. For each such issue, we describe more extensively only one paper to exemplify the approach, and we briefly mention other papers highlighting only their key conceptual differences concerning the former. Finally, in section \ref{sec:challenges}, we highlight critical open challenges, emphasizing that, in most cases, they could be addressed appropriately through a more structured approach to model social aspects of the users' behaviour as they emerge by analyzing data available on the considered Online Social platform. These challenges arise mainly from the nature of social data, its volume, dynamics, and the need for the user to be more aware of the personal data used.

This work aims to enrich the Social Search research area by offering a new schema of the types of social systems existing today and the different methodologies used to make these systems more tailored to users. After identifying the macro-areas of Social Search, we selected and analyzed in detail the methods proposed over the last few years, highlighting the most efficient ones. Moreover, this analysis underlines open challenges that could be addressed by exploiting user models and social information in a different way. The goal is to introduce the reader to the Social Search research area by providing an easy-to-follow overview of its up-to-date state-of-the-art.

\section{Definitions and Related Surveys}
\label{sec:related}
The initial concepts regarding Social Search relate to the modern idea of Social Web Search, which aims to improve the IR processes of search engines leveraging social information. In the absence of popular social networks, the first approach was to consider the history of searches as a source of additional information for future searches via automatic query terms expansion \cite{fitzpatrick1997automatic}.
According to \cite{DBLP:journals/corr/cs-MA-9809025}, Social Search should contrast the simple surfing, where users follow hyperlinks until they find the information. 
Considering the Web as modeled by a graph that connects information in a highly complex way, its huge size presents users with two difficulties: deciding which site to choose and judging a priori the informative value of the returned results. 
Following a Social Search approach, users should first use information from other users' experiences to start with a site and further exploit it to refine the information obtained through regular surfing. Subsequently, the definition of Social Search was transformed by the introduction and popularity of the first social networks. In~ \cite{freyne2004experiment}, Social Search has been defined as \textit{"an approach to Web search that tempts to offer communities of like-minded individuals more targeted search services, based on the search behavior of their peers, bringing together ideas from Web search, social networking and personalization."} 
Social Search becomes a combination between social networking and Web search, exploiting the information about the communities that social networks allow to create. In this way, search engines have become more sensitive to the needs and preferences of specific communities of users. Personalizing search results based on like-minded individuals has also been known as \textit{Social Navigation}.
According to \cite{brusilovsky2005comprehensive}, Social Navigation aims at taking advantage of the human tendency to follow other people with similar interests and repeat their steps.
The authors in \cite{freyne2007collecting} combine Social Navigation and Social Search techniques to take full advantage of the collaboration both can offer. With the former, it is possible to leverage the past browsing behavior of users to guide other users to relevant information; the latter aims to use the search user and users' community patterns to adapt the result list to the needs and preferences of a specific community.

Another evolution in the concept of Social Search emerged with the advent of Web 2.0. Because in Web 2.0 everyone can create content, users have created annotations for Web pages at a surprising speed. An example of a popular bookmarking site was Delicious, with millions of registered users. Annotations represent new information to be exploited to improve IR processes. In \cite{bao2007optimizing}, the authors integrate social annotations into the concept of Social Search by proposing two new algorithms based on PageRank: SocialSimRank and SocialPageRank.
Accordingly, the concept of Social Search has been expanded. In \cite{evans2008towards}, the authors define Social Search as \textit{"an umbrella term used to describe search acts that make use of social interactions with others. These interactions may be explicit or implicit, co-located or remote, synchronous or asynchronous."}
Following such a broad definition, in \cite{carmel2009personalized}, the authors study the effectiveness of different social network types for personalization: (i) familiarity-based network, (ii) similarity-based network, and (iii) overall network. They use the notion of Social Search to describe the search process over Web 2.0 \emph{social data} representing different entities and their interrelations.

A few studies in the literature have proposed categorization of Social Search research. Authors of \cite{mcdonnell2011social} rely on the simple notion that \textit{"Social Search refers to the use of social media to aid finding information on the Internet,"} identifying five crucial dimensions for understanding Social Search:
\begin{enumerate}
    \item \textbf{Collaboration - asynchronous vs. synchronous}: in an asynchronous scenario, the data coming from social media are used to improve the Web search without any interaction between users. Vice versa, the synchronous scenario involves user interactions during the search process;
    \item \textbf{Collaboration - implicit vs. explicit}: the tacit collaboration, as for the asynchronous scenario, is where the search system implicitly integrates social information obtained from users. When people interact during the collaboration, we talk about explicit collaboration;
    \item \textbf{Search target - finding people vs. finding resources}: Social Search is often seen as searching for people, while Social Search systems offer a means to also search for resources by considering social information;
    \item \textbf{Finding - search vs. discovery}: Social Search includes traditional information retrieval mechanisms for searching by re-ranking search results based on social information. However, it also offers the possibility to discover information and resources. For example, on a social bookmarking site, the user can discover new websites that generally are not returned by the system during a search process;
    \item \textbf{Search results - sense-making vs. content selection}: most of the time, people do not fully understand why they get specific results. Social Search can improve sense-making and content selection, providing contextual information.
\end{enumerate}

Overall, the above five dimensions of Social Search are strictly interconnected and work together to enhance the search process by providing users with relevant information and context to understand search results. 
The Social Search process starts considering the search target, which can be finding people or resources. The finding process can be distinguished into search or discovery. During the finding process, people establish some kind of collaboration that can be asynchronous or synchronous and implicit or explicit. Finally, the Social Search process can improve the sense-making of the results or help users in selecting the right content.
By understanding these dimensions, researchers and developers can design and implement effective Social Search systems that meet users' needs.

Finally, in \cite{bouadjenek2016social}, the authors propose a taxonomy of  social IR  defined as \textit{"the process of leveraging social information (both social relationships and the social content), to perform an IR task with the objective of better satisfying the users' information needs."} The taxonomy includes Social Web Search, Social Search, and Social Recommendation. According to the authors, Social Web Search deals with techniques that can improve IR processes by using social information. At the same time, Social Search is \textit{"the process of finding information only with the assistance of social entities, by considering the interactions or contributions of users."} In this way, Social Search is associated with platforms specifically designated for managing social data, such as Facebook or Twitter. The authors further divide Social Search into three categories: Social Question\&Answering (Q\&A), Social Content Search, and Social Collaborative Search.
\section{Paper Selection Methodology}
\label{sec:methodology}

Before presenting the literature according to the taxonomy of Figure~\ref{fig:taxonomy}, we briefly describe the methodology used to identify relevant works on Social Search. Table~\ref{tab:selection} reports summary data. We started collecting documents following the Social Search taxonomy proposed in \cite{bouadjenek2016social}. For each topic (first column in Table~\ref{tab:selection}), we first retrieved papers from Google Scholar using the keywords specified in Table~\ref{tab:selection}. We filtered results based on the year of publication: we chose to consider the works published from 2018 onward to focus on the most recent results in this area and to offer readers up-to-date knowledge. We selected the first 20 results, then analyzed these papers and kept only the most relevant ones according to the topics. Using these papers as initial seeds, we explored the literature using  ConnectedPapers\footnote{\url{https://www.connectedpapers.com}}. ConnectedPaper is a visual tool to find papers relevant to a specific field.
Given a queried paper, Connected Papers explores the Semantic Scholar Paper Corpus to identify papers with the strongest connection: this is done by building a directed graph where nodes are related papers and edges are weighted according to a similarity measure based on Co-citation and Bibliographic Coupling  \cite{kessler1963bibliographic}.
ConnectedPaper returned 41 related papers for each one of the queried works. We removed duplicated papers, and, as done for Google Scholar, we only considered the papers from 2018 onwards. The total number of papers collected through this process is reported in the "CP" (Collected Papers) column in Table ~\ref{tab:selection}. These papers also include articles cited by the most relevant works (according to our opinion) and those citing each work. The set of collected papers is the starting point for our detailed analysis. Specifically, after reading those papers, we kept only a subset of them, whose number is shown in the "SP" (Selected Papers) column of Table~\ref{tab:selection}. Although collecting papers through keywords or tools (such as Connected Papers) produced many articles, only a small percentage of them used community information to improve information retrieval within social platforms: this explains the substantial decrease in the number of selected papers.

Further scrutiny of those papers leads us to report a subset of them in the survey, as indicated in the "RP" (Reported Papers) column of Table~\ref{tab:selection}. The main criteria we used for filtering those papers are (i) the impact of the work in the community (by analyzing citation scores), (ii) the originality of the proposed approach concerning the literature, (iii) the quality of the publication venue; (iv) the level of innovation on the topic. The most crucial factor that influenced the exclusion or consideration of a paper was the use of social data: we focused more on works that integrated data relating to social relationships between users. 
The number of citations and the quality of the venue were secondary factors. Given that the number of citations reflects the article's impact on the scientific community, we preferred studies that attracted citations to those not cited. Finally, we evaluated the originality of the approach and the novelty brought to this research field.

\begin{table}[ht!]
\small
\centering
\begin{tabularx}{\columnwidth}{|>{\hsize=.35\hsize}X|>{\raggedright}X|s|s|s|}
\hline 
\centering \textbf{Topic} & \centering \textbf{Keywords} & \textbf{CP} & \textbf{SP} & \textbf{RP} \\ \hline
\textbf{Expert Finding} & Expert Finding in Community Question\&Answering, Expert Representation, Expert Recommendation, Question Routing, User Expertise, Expert Ranking & 240 & 23 & 15 \\ \hline
\textbf{Question Retrieval} & Question Retrieval in Community Question\&Answering, Duplicate Question Detection & 190 & 27 & 11 \\ \hline
\nohyphens{\textbf{Answer Selection}} & Answer Selection in Community Question\&Answering, Answer Ranking & 105 & 29 & 5 \\ \hline
\textbf{Content Search} & Social Content Search, Social Platforms, Social Information Retrieval, Content Search in Microblogs, Query Suggestion, Query Expansion, Query Auto-Completion & 240 & 29 & 10 \\ \hline
\textbf{Social Collaborative Search} & Collaborative Search System, Collaborative Web Searching, Collaborative Information Seeking & 208 & 22 & 10 \\ \hline
\caption{Keywords and the number of collected (CP), selected (SP), and reported (RP) papers for each topic.}
\label{tab:selection}
\end{tabularx}
\end{table}
\section{Expert Finding in Social Q\&A}
\label{sec:expert_finding}

Community Question\&Answering sites host domain-specific communities where people share their experiences to help other users through point-to-point interactions. However, the engagement of expert users in the community is complex, and not receiving any answer can create a sense of frustration in a user posting a question. A question can fall through unanswered due to several factors, but one of the main reasons behind this behaviour is that the system needs to propose the question to the right expert users. This consequence is reasonable, considering the large number of questions submitted to these sites daily. The task of adequately matching user questions and expert users is commonly called the \textit{Expert Finding Task}. Previous surveys divide the existing literature according to the specific methodology to address the above problems. In \cite{wang2018survey}, the authors classify the studies into eight categories: simple models, language models, topic models, network-based methods, classification methods, expertise probabilistic methods, collaborative filtering methods, and hybrid methods. Instead, in \cite{yuan2020expert}, the authors classify the studies into matrix factorization-based models, gradient boosting tree-based models, deep learning-based models, and ranking-based models. We formalize the problem as follows:
\begin{definition}
Given a question $\bm{q}$ posted by a user  $\bm{u}$, a set of $\bm{n}$ experts $\bm{E= \{ e_1,e_2,...,e_n \} }$, and a positive integer $\bm{k \leq n}$ find the best ranked list of $\bm{k}$ experts who are more likely to answer correctly question $\bm{q}$.
\end{definition}
This task involves mainly two phases:
\begin{enumerate}
    \item \textbf{Expert and Question representation}: the representation of experts should capture their \emph{knowledge} and \emph{interests}. Usually, past questions and answers are the information source considered to extract topics addressed by the experts and their interests. The \emph{reputation score} is also essential as it captures the community's view of the expert quality. Similar to the reputation score, some studies compute an \emph{authority score} that involves the construction of an interaction network on which researchers apply link analysis techniques. The final representation model should cover expertise, reputation, user activity, and the experts' likelihood of answering the question. Analogously, question representation models the topic involved in the question and, in some cases, the interest and profile of the asker.
    \item \textbf{Expert-Question similarity and ranking}: a score for each expert-question pair is computed and used to rank the experts based on some measure of interest, e.g., the expert's likelihood to answer the given question.
\end{enumerate}
Unlike these prior works, we classify existing literature according to the two phases described above, representing a more comprehensive way of characterizing the topic. Therefore, the following subsections present two required modules in any workable Q\&A system.

\subsection{Expert and Question Representation}
\label{subsec:exp_quest_rep}
The expert and question representation phase aggregates all the user aspects (i.e., activities and reputation) to capture their expertise and represent queries, permitting an effective matching between them. The first group of work focuses on encoding information about the user's profile, generally described as features fed into conventional topic modelling or word embedding tools.
In \cite{roy2018finding_ex}, authors propose a method to detect a group of possible answerers based on their history of questions and answers. The authors use features from the user's profile and the community to model the user. Precisely, they formalize a measure called \emph{TagTrueness (TT)} that captures the tendency of a user to answer questions with the tags in her profile:
\[TT = \dfrac{\text{Total Number of Accepted Answers of Profile Tag}}{\text{Total Number of Accepted Answers}}\]
Given a question, users belong in the set of possible experts if they have given at least one accepted answer to a question with the same tags. Each user is associated with a score computed considering TT, the Number of Tags in the Profile (NPT), the Number of Accepted Answers (NAA), the Average Number of Accepted Answers with Tags in Common between the question and profile (AACT), and a Recent Activity Measure (RAM). 
The idea is that users who have \textit{recently} answered questions with tags in common with the question should have a higher score than users who have responded to similar questions in the past. The score is computed as follows:
\begin{equation}
\label{eq:roy2018finding_ex}
    Score=\mbox{TT} \cdot log(\mbox{NPT}+10) \cdot log(\mbox{NAA}+10) \cdot \mbox{AACT} \cdot \mbox{RAM}
\end{equation}

In the same group (user's features modelling), \cite{mumtaz2019expert2vec} proposes a framework combining textual information, community feedback, and user interactions. Authors use word embeddings to represent users' answers and degrees of expertise to capture the semantic relationship between the question and user expertise. The proposed approach overcomes the well-known limitations of Latent Dirichlet Allocation (LDA) \cite{blei2003latent} for topic modelling. 
On the other hand, the original approach of~\cite{tondulkar2018get} is to define embeddings that favour experts who provide few but high-quality answers to complex questions rather than experts who offer solutions that do not satisfy the user or answer only accessible questions. 
Finally, \cite{fu2020recurrent} accounts that sometimes it is possible to learn from past answers/questions which topic a user could face with some expertise. For example, suppose it is possible to deduce from the user information that she is a researcher: the system could forward the question "How can I write a scientific paper?" even if she had never answered this topic. To this aim, the authors define a new approach based on the Recurrent Memory Reasoning Network (RMRN), composed of different reasoning memory cells that implement attention mechanisms to focus on various aspects of the question and retrieve relevant information from the user's history. 
Similarly to \cite{mumtaz2019expert2vec, fu2020recurrent}, word embeddings are also used in \cite{fukui2019suggesting} to enrich the new question with tags semantically similar to those already present to include also experts that, considering only the original tags, would not be considered.

Even if the studies above consider some community-based features, they do not explicitly consider the social network between answerers and askers. Expert Finding approaches based primarily on text, such as those mentioned above, aim to quantify the user's knowledge about a topic based on the questions/answers asked/given in the past. On the other hand, network-based approaches aim to determine the user's authority within the network by considering past activities independently of the new question. In \cite{kundu2019formulation}, authors introduce three new concepts: question hardness, question answerer association, and answerer's performance-enhanced authority. These three concepts are based on three assumptions, respectively: (i) complicated questions require a good knowledge of the topic and help identify the experts; (ii) an answerer is very familiar with a given topic if she has answered multiple questions on that topic; (iii) good answers usually get many votes, so merging information from the network can help. The framework includes the Knowledge Analyzer component aimed to estimate an answerer's knowledge of a topic by combining the output of the following three components:
\begin{itemize}
    \item \textbf{Question similarity finder}: it computes the similarity between the new question and archived questions by treating them as documents and using Query Likelihood Language (QLL) \cite{ponte1998language};
    \item \textbf{Question answerer association provider}: it establishes the relationship between a question and each answerer using conditional probability;
    \item \textbf{Question hardness estimator}: it computes the complexity of the question using features such as Time Response and Number of Answerers. The assumption is that the delay in receiving the first answer to the question and a low number of replies are indices of the complexity of the question.
\end{itemize}
The Authority Analyzer is network-based and consists of two parts: Network Builder and Authority Estimator. The first deals with building a Competition Based Expertise Network (CBEN) \cite{aslay2013competition} where each node is an answerer, and there is a direct link from $a_i$ to $a_j$ if both users answer the same question and $a_j$ is the best answerer of the question among the ones available. The weight of the link $v_{i,j}$ measures how much $a_j$ is better than $a_i$ by considering the number of votes received, and it is computed as follows:
\begin{equation}
     v_{i,j}=\sum\limits_{q \in Q} \delta_i^j(q) \cdot \sigma(a_i,q)
\end{equation}
where $\delta_i^j(q)$ is equal to $1$ if $a_i$ and $a_j$ are co-answers of $q$ and $a_j$ is the best answer, 0 otherwise. Instead, the factor $\sigma(a_i,q)$ incorporates into the computation the information about the quality of the answer $a_i$ concerning the question $q$, and it considers the number of votes received by $a_i$ for $q$ and the maximum votes received by any answerer for question $q$. Instead, the Authority Estimation calculates the authorship of the answerers using link analysis techniques such as AuthorRank \cite{liu2005co} and Weighted HITS \cite{li2002improvement}. 

Other variations of this approach have been proposed: \cite{kundu2020preference} presents a Preference-Enhanced Hybrid Expertise Retrieval (PEHER) system with a preferability estimator. The preferability estimator incorporates the preferences of an answerer $a \in A$ by considering two preferences: \emph{intra-profile} and \emph{inter-profile}. The intra-profile ones capture a given user's preferences for specific terms. In contrast, inter-profile preferences capture the features of the particular answerer concerning all other relevant answerers in the system.

\subsection{Expert-Question Similarity and Ranking}
Depending on the models adopted for representing experts and user queries, we have different solutions for matching and ranking experts likely to answer a given question effectively. Most works cited in Section~\ref{subsec:exp_quest_rep} also address this problem. Therefore, we first briefly describe the approach proposed by them. 

In \cite{roy2018finding_ex}, after ranking the experts using the score computed by Eq. \ref{eq:roy2018finding_ex}, the candidates are re-ranked based on their answering behaviour by considering if a user is generally more active during the day or night, during the week, or on the weekend. Given a question and the corresponding posting time, the experts on the question topic are re-ranked by putting in the top positions the experts that usually are active at that specific time of day. Similarly, Kundu et al. \cite{kundu2019finding} rank the experts by combining their activeness over time with other features measuring expertise and answering intensity. Mumtaz et al. \cite{mumtaz2019expert2vec} rank the experts using a weighted combination of the cosine similarity between users' expertise and question representations.
In \cite{tondulkar2018get}, the authors apply the Learning to Rank methods to the expert ranking task. Specifically, they use the LambdaMART algorithm to learn a ranking model trained on the pairs
\textit{(Questions, Users)} modelled with the four categories of features described in the previous section and labelled based on the historical interaction information. For a new question, the candidate experts are ranked by the score predicted by this model. 
In \cite{kundu2019formulation},  the knowledge and authority scores described in the previous section are combined using a Reciprocal Rank Fusion (RRF) technique \cite{zhang2003expansion}.
RRF is also used in \cite{kundu2020preference} to generate the expert list. 

Other works in the literature focus more specifically on the issue of Expert-Question Similarity and Ranking, while they do not deal mainly with the problem considered in Section~\ref{subsec:exp_quest_rep}. Conceptually similar to \cite{kundu2019formulation, kundu2020preference}, in \cite{le2018retrieving}, the authors develop a network-based framework combining four features: (i) the similarity between the question's content and the user profile; (ii) the similarity between question topics and user topics; (iii) the similarity between asker and answerer in the network; (iv) the activity level of the user. 
Authors compute the similarity of the content through cosine similarity, incorporating the Term Frequency - Inverse Document Frequency (TF-IDF) to weigh the words differently. Instead, they first use LDA to extract topics and then the TF-IDF again to calculate the similarity between topics. 
For the similarity between the answerer and asker within the social network, they build a graph with an edge between two user nodes $u_1$ and $u_2$ if the user $u_2$ has answered a question of user $u_1$. Authors compute the proximity of two nodes by performing Random Walks with Restart (RWR): precisely, given the graph's size, they apply the fast RWR algorithm proposed in \cite{tong2006fast}, whose intuition is to divide the graph into small communities connected by bridge edges and combine the RWR scores computed on these communities. Finally, authors use metrics available on the platforms for the answerer's activity level, such as the reputation score or the number of awards received. Lastly, the authors compute the final score by combining these four factors weighted by parameters learned using the history of the response activity. The study of \cite{faisal2019expert} also considers the answers' quality and consistency. The authors propose a model based on an adaptation of the bibliometric g-index to measure the consistency of the user in providing high-quality answers. They also compute another score that measures the user's reputation (REP-FS) using voter reputation, up-vote to down-vote ratio, participant-based reputation, and popular tags. Weighted Exp-PC combines the two scores to compute the final user's expertise.

Another aspect to consider when routing new questions to experts is that, in a cQA site, each user can cover both the asker and answer role simultaneously, which may vary over time.
In \cite{fu2019tracking}, the authors represent the user considering the evolution of her role over time, and they study how much this aspect can improve the question routing process in a cQA site. To track the evolution of user roles, they propose a Time-aware Role Model (TRM) based on LDA to model the latent topical relationship between question content, user, time, and role. 

Moreover, many approaches do not consider the intimacy between the asker and the answerer. In \cite{fu2020user}, the authors consider this factor based on the assumption that an expert has more incentive to answer a question posted by an asker if she is interested in the topic and if there is intimacy between the two. They propose a User Intimacy Model (UIM), i.e., an LDA-style model that incorporates social interactions in modelling and learning the intimacy between the entities on topics.

Finally, one further aspect to consider in the Expert Finding task is the probability that the expert to whom a given question is forwarded will respond. To approach the problem, according to \cite{DBLP:conf/ijcai/Qian0W18}, it is necessary to find a ranking function that quantifies a user's expertise and the likelihood that they will answer the question simultaneously. To this end, the authors exploit the \emph{Social Identity Theory} \cite{turner1986significance, tajfel2004social} to define a graph joining different possible experts who could answer the question.
\section{Question Retrieval in Social Q\&A}
\label{sec:question_retrieval}

Users often post questions that already exist in the system and have been answered. Posting an already-asked question on a cQA site can negatively affect user engagement by not getting any answers from community experts. It also causes an unnecessary increase in the amount of data the system stores. For this reason, cQA sites provide a service whereby, given a user question, they return all past questions, which may be the same or very similar to the user's question. We can formally define the task as the following retrieval task:

\begin{definition}
Given a query $\bm q$, a historical set of already posted and successfully answered questions $\bm{Q = \{q_1,...,q_m\}}$, and a positive integer $\bm k$, return a ranked list of the $\bm{k \leq m}$ historical questions that are the most similar to $\bm q$.
\end{definition}
At this point, the user submitting $q$ is presented with the answers of the $k$ similar past questions and can quickly decide whether the answers satisfy the information need or whether to post the new question. However, the previous task is complex: different users with the same information need can write semantically equivalent questions using completely different sentences from a syntactical point of view. A significant body of works tries to overcome this \textit{semantic gap} problem using NLP techniques and user metadata information.

As highlighted in Figure~\ref{fig:taxonomy}, the problem consists of two logical phases: \textit{Metadata and Question Representation} and \textit{Query-Question Similarity and Ranking}. 
The two phases are conceptually similar to those in the Expert Finding problem but now refer to a different task. In the Metadata and Question Representation phase, the proposed approaches address the problem of representing the questions and which user metadata could improve the performance of the Question Retrieval service offered by the site. Instead, in the Query-Question Similarity and Ranking phase, the proposed approaches develop methodologies to exploit the content of the questions and leverage the user information available to provide accurate matches between a new query and existing questions. We dedicate the first subsection to the Metadata and Question Representation. Instead,  we describe the Query-Question Similarity and Ranking methodologies dividing them into three subsections corresponding to three approaches that, as shown in Figure \ref{fig:taxonomy}, are alternatives to address this problem: (i) Text-based Similarity and Ranking, which only use only textual data; (ii) Social and Text-based Similarity and Ranking, which also introduces social aspects by modelling users and interactions; (iii) Domain Adaptation, where data is adapted or integrated to improve performances.

\subsection{Metadata and Question Representation}

A very popular technique to represent questions and answers relies on Word Embeddings, which transforms a sentence into a fixed-size dense vector, somehow capturing the semantics of the sentence. In \cite{shah-etal-2018-adversarial, prabowo2019duplicate} authors use GloVe, while in \cite{chen2018question,othman2019manhattan,othman2019enhancing} authors use Word2Vec. For example, in \cite{othman2019manhattan}, after the first filtering phase in which authors apply \textit{text cleaning}, \textit{tokenization}, \textit{stopwords removal}, and \textit{stemming}, each question is represented as a list of terms. Each word is mapped into a fixed-length vector using a pre-trained Word2Vec model. Considering that short texts amplify the mismatch problem, the authors propose expanding the query with other terms close in the semantic space created by the Word2Vec model.

An interesting study from the question representation point of view is presented in \cite{wang2018concept}. Authors try to address five problems when representing questions: \textit{synonymy}, \textit{polysemy}, \textit{word order}, \textit{question length}, and \textit{data sparsity}. The framework consists of two steps:
\begin{enumerate}
    \item \textbf{Word and concept learning}: they use a Skip-gram model to learn the embeddings of the words. In this way, the meanings of words are distributed across the dimension of a semantic space. Furthermore, previous works show that conceptual information can handle the polysemous problem: in \cite{cheng2015contextual}, the authors compute the contextual representation of a word by combining its vector and the most context-appropriate concept vector that delivers an unambiguous meaning. For example, considering two queries [python zoo] and [python string], the embedding of the first "python" is obtained by combining the intrinsic vector of "python" and the concept vector of "animal category." In this way, the model is context-dependent. The problem with the above approach is that the objective function locates the right concept and then searches for a word underneath the chosen concept. In \cite{wang2018concept}, the authors add a regularization function to overcome the problem, which aims to leverage the concept information regarding the knowledge base without considering the context. Considering the context-independent relation could avoid the error introduced by choosing the wrong concept vector;
    \item \textbf{Question and answer embedding learning}: to encode concept, syntactic, and word order information, authors propose the High-level feature embedded Convolutional Semantic Model (HCSM). The model assumes that if two questions have similar answers, they could have semantic relations. HCSM is a convolutional architecture composed of wide convolution (without zero-padding) and pooling. The authors design three variants: the first variant emphasizes the relation (i) between concepts, (ii) between words, and (iii) between concepts and words. The second variant emphasizes only the first two relations, (i) and (ii). The last variant combines the word-embedding question matrix and concept-embedding question matrix before inputting them into the network.
\end{enumerate} 


\subsection{Text-based Similarity and Ranking}
\begin{figure}[ht!]
    \centering
    \includegraphics[scale=0.55]{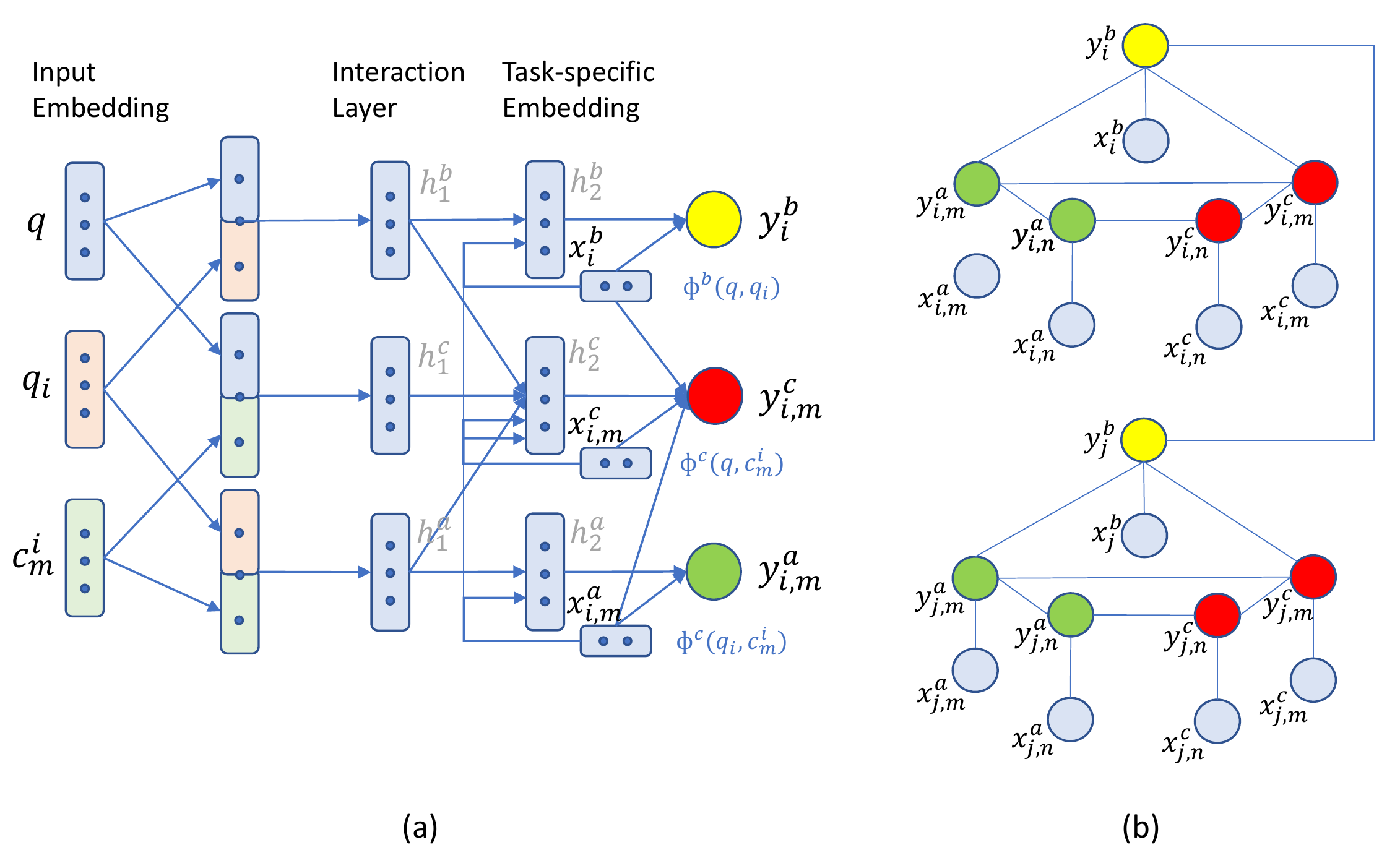}
    \caption{Graphical representation of the cQA framework in  \cite{joty-etal-2018-joint}. On the left (a), we have three feed-forward neural networks to learn task-specific embeddings for the three cQA subtasks. On the right (b), a global conditional random field (CRF) models intra- and inter-subtask dependencies.}
    \label{fig:180908928}
\end{figure}

Joty et al. \cite{joty-etal-2018-joint} provide a remarkably complete approach to addressing this problem. Specifically, the authors observe that some questions may have a lot of answers, some significant and some useless. A long thread of answers for a question can bore the user who decides not to read it and post a duplicate question.
The authors thus try to accomplish the two tasks by introducing a third task: checking whether an answer in a long answer thread is good.
The three sub-tasks are defined as follows: (i) determine whether the $m$-th answer $c^i_m$ in a given thread is good or not for its related question  $q_i$ (task $A$); (ii) calculate whether an archived question  $q_i$ is relevant to the new query $q$ (task $B$); (iii) rank each answer for $q_i$ as relative or non-relative for  $q$ (task $C$). These tasks have dependencies: if an answer $c^i_m$ is good for the archived question $q_i$ and $q_i$ is relative to the query $q$, then $c^i_m$ is a good answer for $q$; if an answer $c^i_m$ is relevant to $q$, then the question $q_i$ to which it refers is also inherent to $q$. Therefore, the idea is to jointly exploit the interactions and dependencies between these tasks to solve the Question Retrieval and Answer Selection tasks. The framework consists of two phases: 
\begin{enumerate}
    \item A Deep Neural Network (DNN) is used to solve the three tasks separately. The subtask-specific hidden layer activations are used as the embedded feature representations for the second step, where a Conditional Random Field (CRF) model performs joint learning with global inference exploiting the dependencies between the three subtasks. As shown in Figure \ref{fig:180908928}(a),  the input for the first step is a tuple $(q, q_i, c^i_m)$ composed of the new query $q$, a historical question $q_i$, and the $m$-th answer $c^i_m$ of the question $q_i$. These elements are converted into fixed-length vectors through syntactic and semantic embeddings and passed to a feed-forward Neural Network intending to learn task-specific embeddings for the three tasks separately. Considering task $A$, shown in the lower part of Figure \ref{fig:180908928}(a), the vectors of answer $c^i_m$ and question $q_i$ are concatenated and passed to the first hidden layer of the DNN. The output is passed to a task-specific hidden layer that combines these signals with a pairwise similarity between $c^i_m$ and $q_i$. The merging between the final hidden-layer output and the pairwise similarity gives the final task-specific embedding. The last layer outputs the prediction variable $y^a_{i,m}$. Similarly, the authors compute the embeddings of tasks $B$ and $C$;
    \item The authors build a large undirected graph consisting of six sub-graphs: $G_a, G_b, G_c$ for the three tasks and $G_{ab}, G_{ac}$, $G_{bc}$ for the three possible interactions between the different tasks. The \emph{intra-subtask} edges are created for the variable of the same sub-task, i.e., $y^b_{i}$ and $y^b_{j}$. Instead, regarding \emph{across-subtask} edges, authors exploit three different types of connection: (i) null or no connection, (ii) 1:1 connection between tasks $A$ and $C$, i.e., $y^a_{i,m}$ and $y^c_{i,m}$, and (iii) M:1 connection to $B$, where all the nodes of subtasks $A$ and $C$ are connected to the thread-level $B$ node. Each node $u$ is associated with its embedding vector $x_u$ and its output variable $y_u$. Each edge that links the two nodes $(u,v)$ is associated with an input-feature vector $\mu(x_u,x_v)$ composed of node-level features, which are (i) comment features and (ii) meta-features. Dependencies between the output variables are modelled by learning node and edge factor functions that jointly optimize a global performance criterion using the CRF model \cite{murphy2012machine}.
\end{enumerate}
The results of the experiments show that the DNN alone achieves good results. Still, the CRF model allows a global inference on a graph structure, considering the dependencies between the different tasks.

Other approaches provide less rich but noticeable solutions for the Similarity and Raking task. The authors of \cite{zhang2018unsupervised} propose an unsupervised framework to compute the similarity between two questions, called Reduced Attentive Matching Network (RAMN). It is based on an attention autoencoder to semantically represent the question, pre-trained using a large amount of unlabeled data, and an attention mechanism to focus on the different parts of the sequence of terms. A different approach to capture the similarity between questions is designed in \cite {zhang2018related}, where authors use a Convolutional Neural Network (CNN) whose input is a set of six matrices deriving from different measurements of similarity between question pairs. Four matrices are vector distances, and two are interaction measurements. These matrices are the input of a Deep CNN with a last fully connected layer which outputs the probability of the question pair being duplicated or related.

Finally, \cite{prabowo2019duplicate} uses a Siamese NN to address the problem. Two questions, $q_1$ and $q_2$, are passed as input to two CNNs to capture their most meaningful features, $\hat{q_1}$ and $\hat{q_2}$. Two additional operations are then performed on  $\hat{q_1}$ and $\hat{q_2}$: (i) matrix multiplication to compute the similarity and (ii) matrix reduction to compute the closeness between the two questions. The result of these two operations is combined with $\hat{q_1}$ and $\hat{q_2}$ and given in input to a Multi Perceptron Layer (MLP) that outputs the duplicate and non-duplicate label.

\subsection{Social and Text-based Similarity and Ranking}

\begin{figure}[!ht]
    \centering
    \includegraphics[scale=0.40]{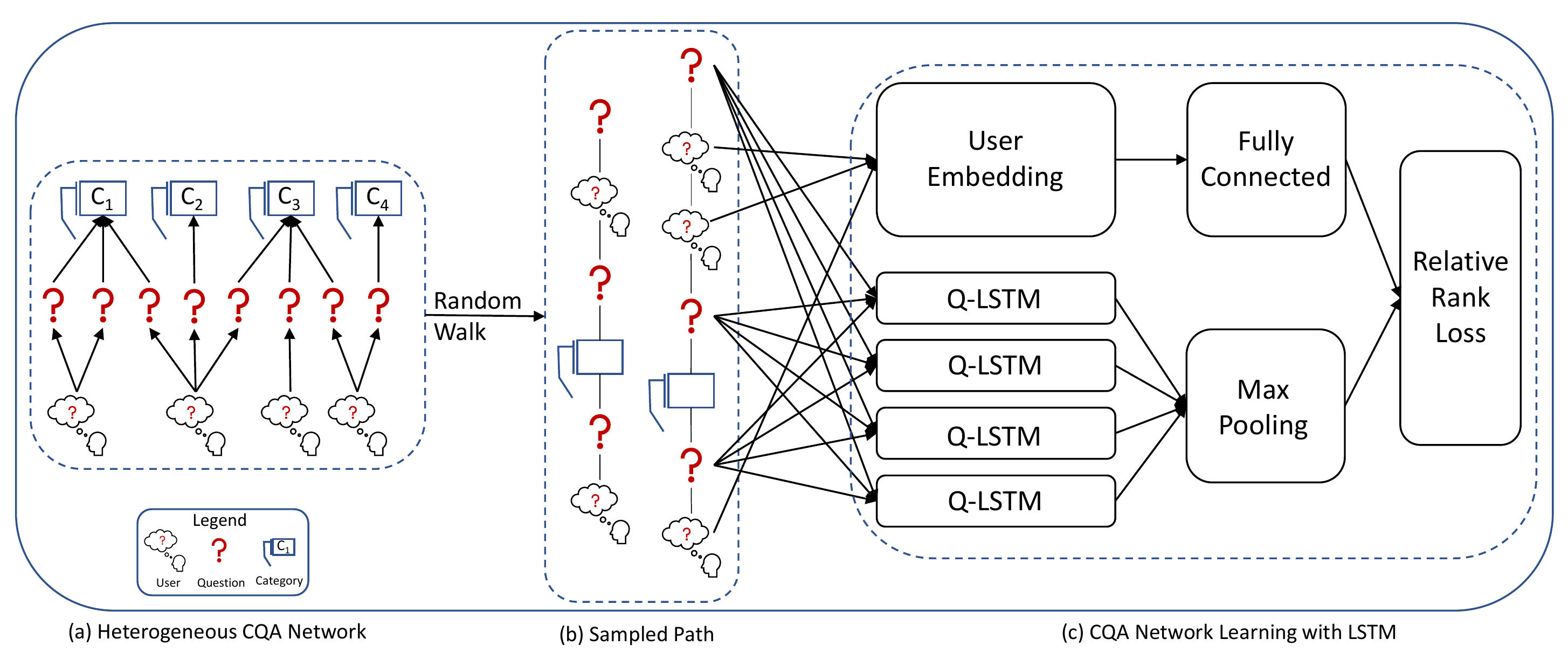}
    \caption{The overview of heterogeneous network ranking learning. (a) The heterogeneous CQA network is constructed by integrating questions content, corresponding categories, and social network relations. (b) A deep random walker walks on the heterogeneous CQA networks to sample the data paths. (c) The questions, askers and categories are encoded into fixed feature vectors and used to train the model for scoring the questions \cite{chen2018question}.}
    \label{fig:chen2018question}
\end{figure}

In \cite{chen2018question}, the authors propose a framework that simultaneously learns the content of questions, their category, and the information in the users' social network to represent the question's intent richly. The framework is illustrated in Figure~\ref{fig:chen2018question}. The assumptions under this approach are two: (i) a user tends to ask questions similar to friends and colleagues because it is likely that they have common interests, and (ii) questions in the same category are more similar than those belonging to different categories. 
Considering the above observations, the authors build a heterogeneous graph composed of three nodes: users, questions, and categories.
A question node is connected to its category node and the user who posted it. This graph is used to sample paths that include different types of nodes applying deep Random Walks inspired by \cite{perozzi2014deepwalk}. Questions are represented using pre-trained word embeddings and LSTM. User embeddings are instead learned from matrices of social relations $M \in R ^{(m \times m)}$ where entry $s(i, j)=1$ if users $i$-th and $j$-th  are friends. Given a new query $q$ along with asker information at inference time, the proposed approach learns, through this heterogeneous network, the latent representation of the queried question. Specifically, it learns the heterogeneous network of graph $G$ and then embeds the asker $u$ and the queried question $q$ to return the top-$n$ similar questions submitted in the past.

Finally, in \cite{hoogeveen2018detecting}, authors address the problem of detecting \emph{misflagged} duplicate questions, i.e., questions marked erroneously as duplicates. It is a complementary approach to the one described in \cite{chen2018question}, and authors use social information related to users (such as their reputation and the scores of their answers) by defining features that help identify errors in marking duplicates.

\subsection{Domain Adaptation}
Finding duplicate questions is a costly process, which is alleviated when experts manually mark duplicate questions. However, some cQA platforms do not offer this functionality, making the Duplicate Question Retrieval process more intricate. One solution is to use Domain Adaptation from another forum. The authors in \cite{shah-etal-2018-adversarial} propose an approach that uses Adversarial Domain Adaptation to help cQA sites that do not have data on duplicate questions. Authors study when the transfer domain can work and what domains' properties are important for successfully transferring knowledge from the source domain. The model consists of a question encoder, a similarity function, and a domain adaptation classifier. The encoder maps the sequence of tokens representing a question into a dense vector. The cosine similarity function predicts if two question vectors $v_1$ and $v_2$ are duplicates. Finally, the domain adaptation classifier predicts whether a given question is part of the source or target domain. The encoder is trained not only to perform well with source data but also to be incapable of distinguishing between question pairs from the source vs target domain. 
The adversarial component reduces the difference between the target and source distributions. For the domain adaptation component, the authors choose the Wasserstein method \cite{arjovsky2017wasserstein}, which diminishes the Wasserstein distance between the source and target distributions. The experiments use three datasets: StackExchange, SprintFAQ, and Quora. The first two datasets are focused on technical domains, while the Quora dataset covers many different topics. Results show that Domain Adaptation works well when the source and target domain are similar (i.e., with StackExchange and SprintFAQ datasets).

Finally, Domain Adaptation is also used to exploit questions written in different languages concerning the target query. These methods include translating the original question written in L1 to the target language L2 and then using a monolingual Question Retrieval model. A relevant example of this approach is presented in \cite{ruckle2019improved}.
\section{Answer Selection in Social Q\&A}
\label{sec:answer_selection}
Answer Selection is an essential process in cQA as it automatically identifies the most valuable answers for an incoming question and proposes them to the user to increase her engagement and satisfy her information need faster. The two main problems for this task are that: (i) answers are usually noisy, and (ii) it is likely that they address different facets of the associated question, not all relevant to the current information need. The problem can be formalized as follow:
\begin{definition}
Given a new query $\bm q$ and a similar historical question $\bm{\hat{q}}$ along with its thread of $\bm n$ answers $\bm{A=\{a_1, ..., a_n\}}$, the Answer Selection problem asks for selecting the best answer in $\bm A$ relevant for $\bm q$.
\end{definition}
Several state-of-the-art works cast the Answer Selection problem to a classification or a ranking task and face it by considering mainly the text of the candidate's question and answer pairs. Notable examples of this approach, identified as \textit{answer-based methods} in the taxonomy presented in Figure \ref{fig:taxonomy}, are \cite{wen2018hybrid, xie2020attentive}, where semantic self-attention and co-attention mechanisms are exploited to focus on the interaction between questions, answers, and their contexts. In the following, instead, we focus on studies that build a graph to model questions, answers, and user expertise, trying to exploit hidden relations between all the entities involved in the selection process. 

\subsection{Graph-based}
The work in \cite{hu2018attentive} proposes a framework that jointly addresses the \emph{redundancy}, \emph{heterogeneity}, and \emph{multi-modality} of questions and answers of the CQA platforms. Text and images are first converted into feature vectors using word embeddings and the ResNet deep CNN model \cite{he2016deep}, respectively, to consider both textual and visual content.
An attention mechanism is then applied to reduce redundancy by constructing a question-answer attentive interaction matrix that focuses on helpful word-pair interactions. Moreover, the authors build a heterogeneous network in which the vertices are questions, answers, users, and tags. Instead, the edges represent the relations among these entities. A meta-path heterogeneous network embedding algorithm is applied to leverage the social information, which allows the discovery of special neighbours of nodes through specific meta-path schemes. For example, considering the schema \textit{"answer-user-answer-question-tag,"}, the last vertex \textit{tag} could be viewed as a special neighbour of the first answer node in the schema. The aim is to learn representations $\overrightarrow{u}_q$ and $\overrightarrow{u}_a$ of questions and answers that consider indirect relations among question, answer, user and tags. Finally, based on the idea that the similarity between word-pair can contribute differently to the final score in a different context, the authors incorporate context information in the similarity computation. 

\begin{figure}[ht!]
    \centering
    \includegraphics[scale=0.5]{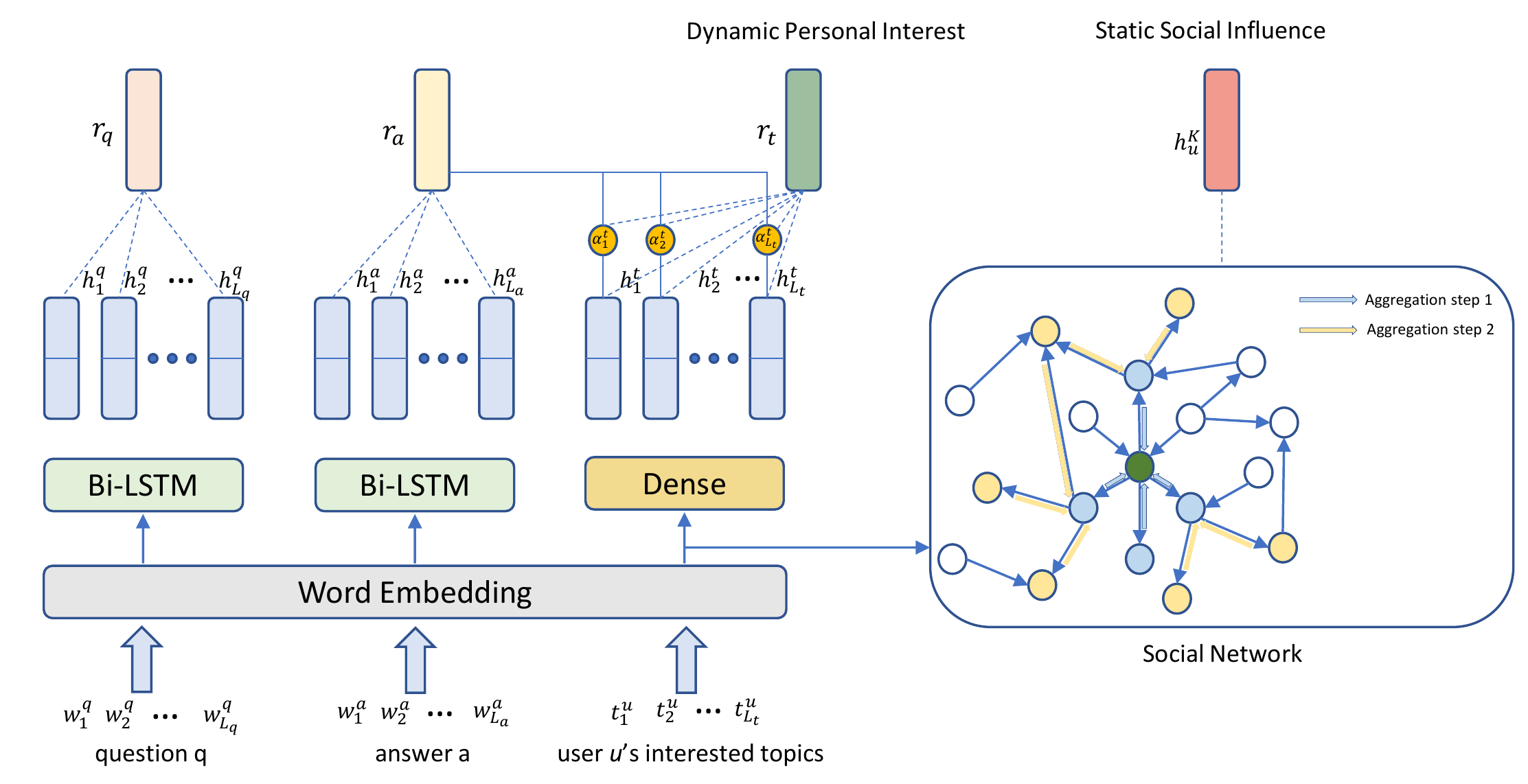}
    \caption{Overview of the proposed SCAD model. It encodes the user expertise by jointly considering the dynamic personal interest and the static social influence. The aggregation process of the social influence takes K = 2 as an example \cite{liu2021toward}.}
    \label{fig:liu2021toward}
\end{figure}

Liu et al. \cite{liu2021toward} observe that modelling user expertise only through historical answers, as in \cite{wen2018hybrid, xie2020attentive} works well only if sufficient data is available. They thus investigate the impact of \textit{static} and \textit{dynamic} social influence on the answer selection problem. They claim that a form of static influence is applied indirectly by other experts or users a person follows in the community: users with a passion for a field will follow experts in that field and read their posts/answers to improve their knowledge of the field. Instead, the dynamic influence refers to dynamic personal interests: users can have different interests, but only a few are activated when they face a new question. Considering all the user interests as equally important can thus negatively affect expertise modelling. Relations in cQA sites are modelled as a directed graph where the nodes are users, and there is an edge from $u_1$ to $u_2$ if $u_1$ follows $u_2$. The framework is illustrated in Figure \ref{fig:liu2021toward}. The authors compute social influence in two phases. During the first phase, the user embedding is the concatenation of the user interest feature vector $x_u$ (obtained using the word embedding of explicit user topics) and a randomly initialized latent vector $p_u$. 
The second phase updates the embedding with $k$ aggregation steps by incorporating, at each stage, the information of the user's neighbours nodes. 
Moreover, the authors model user interest dynamics through a context-based dynamic representation. The embeddings of topics representing user interests are stacked, thus composing a topic embedding matrix. They learn a hidden representation of each topic through a dense layer. Assuming that the answer given by the user is a personal expression containing information about the topics activated at that moment, the authors compute the weights for each topic using the specific answer by a multi-layer perceptron attention mechanism. Instead, personal interests consider the weighted sum of the interesting topics and the global user interest obtained by applying the mean-pooling overall topic representations. The representation $u$ of user context is finally obtained by concatenating social influence and personal interest information. Answer selection is computed by a matching layer that takes as input a triple $(q, a, u)$ and outputs the matching score for ranking the candidate answers. 

Similar to \cite{liu2021toward}, the authors of \cite{zhang2021graph} also remark that users have many interests, but only a few are relevant to a specific question. They address the answer selection process by learning a Graph Neural Network (GNN) from a graph modelling the relations among questions, answers, and terms, weighted with TF-IDF values to capture their different importance for answers and questions. They introduce a novel GTAN (Graph-based Tri-Attention Network) model to encode answer correlation into their representations. Based on the GNN representation, target-aware answerer representation, answer-specific question representation, and context-aware answer representation are computed through attention mechanisms. These three representations are used to measure the final score for the answers. The approach outperforms the one used in \cite{xie2020attentive}, demonstrating the possibility of explicitly encoding answer correlations by GNN modelling and learning answer-specific question representations and target-aware respondent representations.

\section{Social Content Search}
\label{sec:content_search}

Social Content Search refers to the process of retrieving relevant content on social platforms by exploiting information such as clicks, social annotations, social interactions, and personal interests. 
We broadly divide this body of works into two classes presented in Sections~\ref{sub:social-search-no-query-mod} and \ref{sub:social-search-qe-qs}, respectively. As shown in Figure \ref{fig:taxonomy}, the two classes are alternatives for facing the Social Content Search task. Approaches in the first class do not modify the issued query and use social information to determine the relevance of the content that possibly answers the query. Instead, approaches in the second class manipulate the given query, exploiting social information sometimes.

\subsection{Social Content Search without Query Modification}
\label{sub:social-search-no-query-mod}

Most of the time, Content Search uses social information to define specific features in a more complex representation of users, which also involves representations of the content they have generated. In this sense, social information should be defined as any information about the interaction between users or between users and the content they provide (e.g., likes, retweets) available on the social platform.

The work in~\cite{albanese2021improved} is a relevant example in this class. Content modelling always remains essential for retrieval, and, as discussed above, LDA  is widely used in this area. The LDA approach has limitations with short texts and sparse data. Sparsity is observed mainly on social platforms like Twitter, given the limits imposed on the length of posts. To overcome the problem, in \cite{albanese2021improved}, the authors propose a new pooling scheme for Twitter topic modelling that groups tweets based on authors who belong to the same community. The method is called Community Pooling and makes use of a social graph. The nodes in this graph are users, and an edge connects two users based on their retweet behaviour. Authors extract communities using the Louvain method for community detection \cite{blondel2008fast}. The authors associate each community with a document containing all author's tweets who belong to the community. This way, the number of documents is small, but they have enough words for the LDA algorithm to work well.
The method is compared with baselines that group tweets by author, hashtag, or thread of replies under a tweet. The methodology works well for the document retrieval task, where, given a query, the model returns the most similar tweets based on their LDA topic similarity.

Xu et al. \cite{xu2020integrating} use a conceptually similar approach tailored to platforms allowing for the annotation of documents. Specifically, each document is associated with a "document" vector, which is the outcome of an LDA to extract its topic distribution, and a "social" vector, which represents the distribution of topics for users who annotated the document.
Authors of~\cite{xia2020relevance} focus on Twitter and design a system whereby a range of information related to tweets is used for a complex ranking process to define the relevance of tweets. Specifically, features considered include (i) static features (e.g., the creation time, influence of author); (ii) real-time features (e.g., number of social actions received); (iii) query features (e.g., query terms and number of clicks for query results); (iv) personalization features (e.g., the follower-followee graph). It also includes specific information related to the platform's social features, used to identify relevant tweets given a particular query.

Similarly, \cite{huang2020embedding} focuses on Facebook instead of Twitter. The goal is to define a proper embedding of the user issuing the query and the possible documents to be provided. In addition to textual features, the embedding includes social features extracted from the interaction between users and between users and content considered for the query.

\subsection{Query Expansion and Suggestion}
\label{sub:social-search-qe-qs}
To improve the performance of the retrieval process and increase the number of relevant documents returned as output, one of the most common techniques is Query Expansion (QE), which adds new terms to the original query. Formally:
\begin{definition}
Given a query $\bm{q={t_1, ..., t_n}}$ composed of $\bm n$ terms and a positive integer $\bm k$, the Query Expansion (QE) approach adds $\bm k$ terms to $\bm q$ that are likely to occur in relevant documents thus formulating an enriched query $\bm{q*={t_1, ..., t_n, t_{n+1}, ..., t_{n+k}}}$ 
\end{definition}
In literature, many works perform QE by relying on document content. Instead, the expansion of terms also should include user interests and the context
in a social context. An interesting study is the one of Bouadjenek et al. \cite{bouadjenek2019personalized}, where the authors propose an approach for selecting terms for query expansion that is based on (i) the semantic similarity between a query word and a candidate term, (ii) the social closeness between the query and the user (how interesting a query term can be for the user), and (iii) a strategy for expanding user queries online. The approach consists of three phases: (i) select candidate terms that are similar and related to the query based on their co-occurrences with resources and users; (ii) profile the issuer user based on interests and activities; (iii) expand the query considering both the semantic similarity and the user's profile. The method is based on \textit{Folksonomies}, whose definition is the following:
\begin{definition}
Let $\bm{U, T, R}$ be the set of Users, Tags, and Resources, respectively. A folksonomy $\bm{\mathscr F (U, T, R)}$ is a subset of the cartesian product $\bm{U \times T \times R}$ such that every triple $\bm{(u, t, r) \in \mathscr F}$ is a bookmark, representing a user $\bm{U}$ who used a tag $\bm T$ to annotate a resource $\bm R$. 
\end{definition}
The approach consists of two parts. The offline part deals with building the social graph and converting it into a tag graph of a folksonomy $\mathscr F$, where two tags are connected if they are semantically correlated. In addition, the offline part is also responsible for building and updating user profiles. The online part performs query expansion based on the tag graph and user profile.
There are two approaches to constructing the tag graph, starting from the folksonomy:
\begin{itemize}
\item \textbf{Extracting semantics from resources}: semantically related tags are expected to occur on the same resources. The semantic relationship is computed through similarities that require a reduction in the dimensionality of the tripartite graph $\mathscr F$ in a bipartite graph. Specifically, the authors use a function that performs a projective aggregation on the entire folksonomy $\mathscr F$, resulting in a bipartite \textit{Tag-Resource} graph. Then, a $T_R$ tag graph is extracted in which two tag nodes are connected by a weighted edge according to the Jaccard, Dice, or Overlap similarity metrics;
\item \textbf{Extracting semantics from users}: the same users use related tags to annotate resources. Similar to the previous case, the authors perform a projective aggregation on the resource dimension by creating a \textit{Tag-User} bipartite graph. Hence, they obtain a $T_U$ tags graph using the metrics used to obtain $T_R$.
\end{itemize}
Choosing only one method can lead to a loss of information, so the authors combine the two graphs to get a single $T_{UR}$ tag graph that incorporates all the information. The edge between two tags is weighted using the Weighted Borda Fuse (WBF) technique. As explained below, the authors use the graph to extract terms that are semantically related to a given query term.
As a further refinement, the user's interests should be considered to return a result personalized for each user. In the context of folksonomies, a user annotates a resource using tags that summarize her understanding of that resource. These tags are a good summary of users' interests. Therefore, each user is represented as a set of tags, and their weight is computed by adapting the TF-IDF measure based on the user's profile.
In the online part, the authors consider the similarity between a query term $t_i$ and a potential candidate term $t_j$ for query expansion based on tag graph and user interests. The first step is to extract the user profile, given a query and its issuer. Then, the authors extract all adjacent tags for each query term using the $T_{UR}$ graph. For each candidate tag $t_j$, they compute a ranking metric that combines (i) the similarity of a query term $t_i$ and the candidate term $t_j$ given by the graph and (ii) the measure of user interest for tag $t_j$ calculated considering the similarity between $t_j$ and all the tags in the user profile.
Finally, they order the terms of the candidates according to this ranking measure.\\
Instead, in \cite{khalifi2020enhancing}, the authors integrate social scores of users into the similarity between query and document, which are based on two criteria: positivity of a user and feedback.

An alternative to QE is Query Suggestion (QS), which is more powerful concerning synonymy and polysemy problems. The QS process offers the user alternative queries, allowing her to choose which one to explore. 
In \cite{zhong2020personalized}, the authors propose an improved Query Suggestion system for Linkedin by simultaneously modelling structured personalized user features and unstructured text data in a sequence-to-sequence model (Seq2Seq).
Differently, Chen et al. \cite{chen2018attention} designed a Hierarchical Neural Query Suggestion system that combines a session-level and a user-level neural network to model a user's short- and long-term search histories.
Finally, in \cite{kazi2020incorporating}, the authors suggest a novel training approach that considers user feedback as additional machine translation.

Auto-completion is another search engine feature that provides possible query completions as the user is typing. An interesting study is the one of Jaech et al. \cite{jaech2018personalized}, where the authors show how an adaptive language model based on a recurrent neural network can generate personalized completions.

\section{Social Collaborative Search}
\label{sec:collaborative_search}
Although the information search process is usually an individual action, it may involve the collaboration of several subjects. In this context, when two or more people collaborate to satisfy an information need, we speak about \textit{Collaborative Search}. According to Golovchinsky et al. \cite{DBLP:journals/corr/abs-0908-0704}, the Collaborative Search can be studied along four dimensions: (i) the intent, which is explicit or implicit; (ii) the mediation, applied through user interfaces or algorithms; (iii) the concurrency, that can be synchronous or asynchronous; (iv) the location, which can be remote or co-located. Afterwards, Morris et al. \cite{morris2013collaborative} suggested two additional dimensions: the role (symmetric or asymmetric) and the medium (Desktop or emerging devices). 
Given the focus of this study, we present the literature according to the following categories: (i) solutions oriented towards specific communities, (ii) general-purpose solutions, and (iii) solutions based on chatbots. The three categories are alternative approaches, as Figure \ref{fig:taxonomy} illustrates.

\subsection{Solutions Oriented Towards Specific Communities}
\label{sub:CIS - specific communities}
With the specific aim of helping the academic world, in \cite{nedumov2019scinoon}, the authors propose SciNoon, a system capable of facilitating the exploratory search process students and researchers perform in their daily work. Unlike traditional search, exploratory search requires a significant effort, mainly to estimate the results' relevance, where a direct collaboration can help make assessment faster and more precise. To this end, SciNoon provides collaborators with a shared workspace of collected articles and a chatbot that allows reporting every activity of team members. 
It integrates Google Scholar suggestions for related papers and queries. It also extracts from retrieved papers keywords related to the user search intent. The information is organized in the workspace according to a graph-based data model: the nodes are articles, authors, searches, and associated results; the edges express the relationship among them. The system provides a content-based recommendation tool that allows users to expand an article node based on citing and cited relationships.

SearchX \cite{putra2018development} is a collaborative search system designed explicitly for large-scale Collaborative Search (CSE) research enabling users to implement and run their CSE experiments. According to \cite{brennan2008coordinating, maican2019study}, we can categorize the features of a collaborative system along three lines: division of labour, sharing of knowledge, and awareness. SearchX provides these features with a group chat for direct communication, a shared workspace, and color-coding.

\subsection{General-purpose Solutions}
\label{sub:CIS - general purpose}
People often perform collaborative search tasks during daily activities, e.g., choosing a restaurant or a product to buy. 
In such a context, group members often have to reach a consensus, which is challenging when the criteria of each member are conflicting. There is a need for communication between the members and awareness of each other's preferences. To this end, in \cite{hong2018collaborative}, the authors propose a component that allows users to reach a consensus through the awareness of preferences, integrating it into a decision-making system. The component, called Collaborative Dynamic Queries (CDQ), acts as a moderator with three roles:
\begin{enumerate}
  \item It provides each member with awareness of the preferences of others without the need for communication;
  \item It shows candidates who match both individual and group preferences, allowing the identification of ideal candidates; 
  \item In case of no agreement, the moderator identifies the sources of disagreement and suggests relaxing the preferences.
\end{enumerate}
The design encourages mutual awareness among members. In particular, each member has a colour associated. These colours will appear both under each filter and next to a candidate. The former allows viewing the preferences of each member, and the latter shows who may or may not agree on the choice of that candidate as the final solution. The authors conduct two user studies and show the moderator's effectiveness in maintaining awareness in the group and facilitating effective and efficient communication by reducing the effort it usually requires.

Furuie et al. \cite{furuie2020showing}, instead, designed a system for personal smartphones that allows comparing web pages shared among users of the same group. This function is automatically invoked differently based on the orientation of the terminal.
In addition to the traditional search process involving keywords, entity-based search has been introduced over the years, where named entities replace keywords. In \cite{andolina2018querytogether}, the authors propose Querytogether, a multi-device search tool in which entities such as keywords, documents, and authors can be used to compose a query or can be shared with other members. 

Considering the asynchronous scenario, looking at the search process and colleagues' results can fill the gap of a user without background; at the same time, her search process and results can help others discover new aspects. In \cite{xu2021search}, building on previous work \cite{xu2018logcanvas, tolmachova2019visualizing}, the authors propose LogCanvas, a prototype of a search tool supporting asynchronous remote collaborative web search. LogCanvas is explicitly designed for search history visualization and, differently from other platforms, aims to reconstruct the semantic relationship among the users' search activities. The search process of a session is represented by a knowledge graph consisting of all queries and essential related concepts and their relationships using Yahoo's Fast Entity Linking toolkit (yahooFEL) \cite{blanco2015fast}. Moreover, LogCanvas groups queries and entities according to the topic by applying a Wikipedia-based categorization method, i.e., TagTheWeb \cite{medeiros2018tagtheweb}.

\subsection{Solutions Based on Chatbots}
\label{sub:CIS - general purpose}
In \cite{avula2017searchbots}, motivated by a study revealing that users use messaging channels external to the collaborative platform to accomplish their search tasks, two different chatbots are incorporated into SLACK: the first bot explicitly requests information, and the second one instead deduces this information through conversations. The study shows that chatbots improve collaborative search thanks to a greater awareness of team members' activities and a greater simplicity in communicating and sharing information and content. 

SECC \cite{zhang2016secc} uses a social engine that allows users to communicate across multiple channels. However, only some people may have the same information need, leading to a cold start problem as the number of people is unsuitable for a collaborative search. To overcome the problem, the same authors in \cite{wang2019scss} propose to insert an auxiliary intelligent robot in a collaborative search platform capable of supporting the communication between the system and a search engine. In this way, if there is only one user in the chat channel, she can continue to search by interacting with the search engine through the mediation of the chatbot. To this goal, the authors apply a Machine Reading Comprehension (MRC) technique to support the conversation between users and the chatbot. The system consists of a Search engine, a Cluster engine, and an Interactive engine. The Search engine provides all the functionalities of search engines, such as query expansion/suggestion and web page ranking. The Cluster engine groups users dynamically based on the keywords of the query they submit. Finally, the Interactive engine is composed of the Social engine and InfoBOT. The former provides communication channels to different user groups to allow collaborative research. The latter collects the conversations, detects the new query, consults the search engine to collect the related documents, and modifies the query by incorporating past discussions. Finally, it passes the documents and the modified query to the MRC model that returns a response.
\section{Open Challenges}
\label{sec:challenges}
This section identifies some open challenges that characterize Social Search tasks. We focus primarily on those where adequate social data processing can improve the effectiveness or efficiency of the Social Search system, as this is the area where we have identified the main gaps and subjects for future research. The current literature does not yet fully exploit the potential of social relationships. Some open challenges listed below can be addressed by integrating community information, while others can lead to a better representation of users and relationships.

\begin{itemize}
    \item \textbf{Modelling Users and their Relations}\\
    The social aspect of Social Search platforms plays a pivotal role. The studies analyzed in this survey show that social features are poorly exploited in some fields (e.g., Question Retrieval). In other areas, such as the Expert Finding task, the social dimension is widely adopted, for instance, by constructing networks to calculate users' authority. However, embedding social information is often not sophisticated and is implemented through a simple combination of features made available by social platforms. One challenge is to model users and explicit/implicit interactions within platforms in a more structured way, leveraging advanced methods proposed, for example, in the social network analysis literature. Integrating advanced interaction models could make the service more user-centric, capturing latent information about relationships. For example, in the Expert Finding task, such a more extended representation of users and their relationships could lead to identifying more efficiently and effectively experts who can provide good answers, thanks to the additional contextual information embedded in the expert finding process.
    
    \item \textbf{Cross-Linking}\\
    The poor use of data about users and their interactions may derive from the need for more availability of such information on a specific social platform. It is known that different social platforms offer different and complementary services, thus collecting content and information that reflect diverse aspects of the user and her social relationships. A way to improve Social Search processes is to exploit several social sources through cross-linking mechanisms of user profiles. Therefore, integrating cross-linking mechanisms would allow for a deeper behavioural and social knowledge of users that the social system can widely exploit to offer a performing service. The impact of cross-linking across different social platforms may be extremely relevant. On the one hand, it could allow Social Search tasks to have rich information about users, irrespective of how active they are on a specific social platform, by complementing sparse information available on many platforms. On the other hand, even for users with rich profiles on a given platform, cross-linking could allow us to obtain a complete profile (either individual or social) of that user, as the behaviour of a given user on different social platforms may be significantly different.

    \item \textbf{Explainability}\\
    One way to make the user more aware of the data used and processed could be to explain why the system provides a specific result or suggestion. A user-friendly explanation of why the user received a particular outcome could make her more aware of the underlying mechanisms. Moreover, this would increase trust, making the users more inclined to share their data and favour personalization. In the analyzed works, the extensive use of black-box models, such as deep neural networks, further intensifies the need for explainability methods \cite{singh2019exs, fernando2019study, zhang2020explainable, singh2020model, zhang2021explain, zhuang2021interpretable, volske2021towards}. In this respect, social information may be used to expose different levels of information to different social groups based on the strength of social ties between users. Explainability techniques would make the platform more transparent regarding processing the user's data. At the same time, the user could feel more stimulated in using the social platform.
    
    \item \textbf{Scalability}\\
    The pervasive use of social platforms has led to an exponential growth of content created and shared by users. Often, the methodologies proposed in state-of-the-art are computationally expensive when applied to large-scale data. Therefore, there is a need to design methodologies that can cope with large data, guaranteeing a good trade-off between effectiveness and efficiency. From this standpoint, social information may be used to build effective filters limiting the amount of data searched to retrieve the few relevant results presented to the users based on their social relationships. The scalability property will allow the system to take advantage of the large amounts of data fully. By limiting algorithms' application to a properly identified subset of data, the system can improve responsiveness by avoiding unnecessary or redundant computations. In this case, the major impact we expect would be to reduce the noise of information used to identify good experts and good answers, with an overall improvement in the effectiveness of the search process.
    
    \item \textbf{Dynamicity}\\
    Users are dynamic entities that continually evolve, changing their interests, behaviours, information, and social interactions. A Social Search system should address the dynamicity of data to offer an effective personalized service focused on the user. Researchers can achieve it by implementing algorithms that can continuously update the indexes with the most recent data while seeking a trade-off between computational costs and the accuracy of the results. Social information may also be used to predict the near-term evolution of social links (e.g., via advanced link prediction algorithms) and configure search processes accordingly. By monitoring the dynamicity of users and their relations, the social platform can guarantee an updated personalized service with the consequence of better user engagement.
    
    \item \textbf{Bias Mitigation}\\
    Social Search systems must learn as much as possible about users' interests and behaviours to provide tailored results that can fulfil the user's expectations. However, the mechanism by which a system provides results based on the user information could introduce a bias creating a situation in which users are presented with homogeneous results from a limited set of similar sources \cite{chen2023bias}. In practice, however, proposing to users diversified results slightly differing from their profile could bring additional value to the service provided by expanding user vision and knowledge of a concept. The introduction and evaluation of diversification methodologies to mitigate social systems' bias have been only partially investigated and deserve further attention. Contrasting the bias effect could foster the diffusion of new sources and sets of expertise, making the platform's results complete and more interesting.
    
    \item \textbf{System Evaluation}\\
    While offline evaluation is a fast and easy way to test systems, it may reflect something other than reality and is often characterized by biases. On the other hand, online evaluation allows for directly evaluating users' degree of engagement and satisfaction. Unfortunately, it remains impractical or reserved for the owners of social platforms. 
    Most of the studies considered in this survey are based on offline evaluation methods. Still, the research community should put additional effort into improving the assessment methodologies by providing standard benchmark datasets and using the same evaluation metrics chosen based on the specific tasks considered. Introducing several up-to-date benchmark datasets can avoid bias and offer researchers a way to compare their studies with social media platform owners as well.
    
    \item \textbf{Cold Start Problem}\\
    The cold start problem \cite{lu2020meta, wei2021contrastive, dong2020mamo, cai2023user} represents one of the limitations that has always existed regarding social platforms. It occurs when more data about a specific user is needed to provide good service (e.g., for new users). The availability of rich information about the communities in the system and the scarce information about the user could mitigate the problem. By solving the cold start problem, the system can guarantee high performance to new users who will feel immediately satisfied, allowing platforms to engage users in a critical stage, i.e., when they enter the system. Cross-linking (discussed above) is a related challenge, which could also help address the cold-start problems, ultimately contributing to an easier engagement of new users.
    
    \item \textbf{Multimodality}\\
    Given the heterogeneity of the information present within the social platforms, it would be convenient to exploit Deep Multimodal Fusion methodologies to take full advantage of the data available \cite{lu2018visual, chen2022hybrid, hu2019scalable, yu2020improving, cui2021rosita, zhu2022multi, zheng2021multimodal, sun2020riva}. In particular, with the fusion of visual media (e.g., images shared by the user), text, and relational information, the user profile could be complete and represent reality as faithfully as possible. Integrating several types of information allows for capturing complementary aspects of the user, allowing the system to increase the performance of the personalization mechanisms.
\end{itemize}



 \bibliographystyle{elsarticle-num} 
 \bibliography{cas-refs}





\end{document}